\documentclass[useAMS, usenatbib, a4paper]{mn2e}
\usepackage{aas_macro, amsmath, amssymb, color, graphicx, multirow, setspace, times}
\begin{document}

\title[The Progenitors of the Milky Way Stellar Halo]{The Progenitors of the Milky Way Stellar Halo: Big Bricks Favoured over Little Bricks}

\author[A. J. Deason et al.]{A. J. Deason\thanks{E-mail: alis@ucolick.org}$^{1,4}$, V. Belokurov$^{2}$, D. R. Weisz$^{3,4}$\\
$^{1}${Department of Astronomy and Astrophysics, University of California Santa Cruz, Santa Cruz, CA 95064, USA}\\
$^{2}${Institute of Astronomy, University of Cambridge, Madingley Road, Cambridge, CB3 0HA, UK}\\
$^{3}${Astronomy Department, University of Washington, Seattle, WA 98195, USA}\\
$^{4}${Hubble Fellow}}
\date{\today}

\pagerange{\pageref{firstpage}--\pageref{lastpage}} \pubyear{2015}

\maketitle

\label{firstpage}

\begin{abstract}

We present a census of blue horizontal branch (BHB) and blue straggler
(BS) stars belonging to dwarf galaxies and globular clusters, and
compare these counts to that of the Milky Way stellar halo. We find,
in agreement with earlier studies, that the ratio of BS-to-BHB stars
in these satellite populations is dependent on stellar mass. Dwarf
galaxies show an increasing BS-to-BHB ratio with luminosity. In
contrast, globular clusters display the reverse trend, with $N_{\rm
  BS}/N_{\rm BHB}$ ($\lesssim 1$) decreasing with luminosity. The
faintest ($L < 10^5 L_\odot$) dwarfs have similar numbers of BS and
BHB stars ($N_{\rm BS}/N_{\rm BHB} \sim 1$), whereas more massive
dwarfs tend to be dominated by BS stars ($N_{\rm BS}/N_{\rm BHB} \sim
2-40$). We find that the BS-to-BHB ratio in the stellar halo is
relatively high ($N_{\rm BS}/N_{\rm BHB} \sim 5-6$), and thus
inconsistent with the low ratios found in both ultra-faint dwarfs and
globular clusters. Our results favour more massive dwarfs as the
dominant ``building blocks'' of the stellar halo, in good agreement
with current predictions from $\Lambda$CDM models.
\end{abstract}


\section{Introduction}
The Milky Way is a cannibal; throughout its lifetime it captures and
destroys smaller dwarf galaxies. The remains of destroyed dwarfs are
splayed out in a diffuse stellar halo, while the dwarfs evading
destruction comprise the satellite population that orbits the
Galaxy. Despite this well-established, generic picture of stellar halo
formation, we have very little understanding of what the building
blocks of the halo actually are; is the halo built up from many small
mass tidbits, or from one (or two) massive dwarf(s)?

The chemical properties of halo stars have often been used to connect
them to their progenitor galaxies. For example, the relation between
[$\alpha$/Fe] and [Fe/H] is an indicator of the rate of
self-enrichment, and therefore can be linked to the host galaxy's
mass. However, the [$\alpha$/Fe] abundances of halo stars appear to
differ significantly from those of the (classical) dwarf galaxy
satellites in the Milky Way (\citealt{tolstoy03}; \citealt{venn04}),
whereby the halo stars are typically more $\alpha$-enhanced at a given
metallicity. Thus, there is little evidence for the accretion of fragments similar to the present-day dwarf spheroidal population.

The mismatch in chemical properties between the bulk of the halo stars
and the stars belonging to dwarf spheroidals can perhaps be reconciled
if the Milky Way halo progenitors are biased towards massive, early
accretion events (\citealt{robertson05}; \citealt{font06}). The
combination of high-mass and early accretion, can lead to abundance
patterns (at least in the [$\alpha$/Fe]-[Fe/H] plane) similar to that
exhibited by the present day halo stars. This scenario has been
supported by recent evidence of a ``break'' in the stellar halo
density profile at $r \sim 25$ kpc (\citealt{deason11};
\citealt{sesar11}). In \cite{deason13}, we argue that this break could
be evidence for a major (relatively early) accretion event. However,
this is not a unique solution; the same broken profile can plausibly
be produced from multiple, but synchronized, lower-mass accretion
events.

A different scenario posits that analogues of the ``ultra-faint''
dwarf galaxies could contribute significantly (at least at the
metal-poor end) to the present-day stellar halo
(e.g. \citealt{frebel10}; \citealt{clementini10}). For example,
\cite{clementini10} argue that the Oosterhoff classification
(\citealt{oosterhoff39}) of RR Lyrae stars in ultra-faint dwarfs is in
better agreement with the stellar halo compared to the more massive
dwarfs\footnote{However, we note that \cite{fiorentino14} recently showed that the period and luminosity amplitudes of RR Lyrae stars in the halo are more consistent with massive dwarfs (such as Sagittarius) than lower-mass dwarfs.}. Thus, an alternative view is that the stellar halo is built-up
from a very large number of puny dwarfs. Finally, bear in mind that the characteristic building blocks of the stellar halo need not be dwarf galaxies. Previous work has argued that a significant fraction of the stellar halo (up to 50\%) could be assembled from destroyed globular clusters (\citealt{carretta10};
\citealt{martell11}).

\begin{figure*}
  \centering
   \includegraphics[width=15cm, height=5cm]{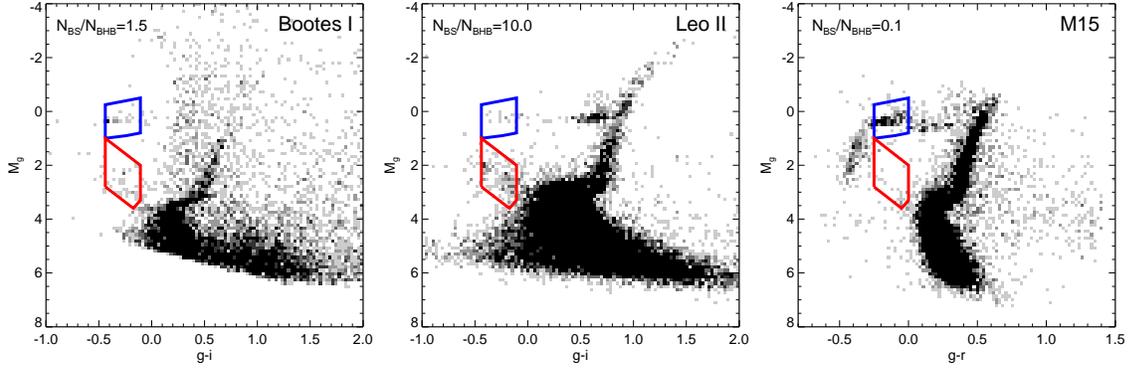}
   \caption[]{\small Three example colour-magnitude diagrams in $gri$ SDSS filters, with original photometry from \cite{belokurov06}, \cite{held05} and \cite{an08}, respectively. The selection of BHB/BS stars are indicated with the blue/red lines, respectively.}
   \label{fig:cmds}
\end{figure*}

Despite the wealth of work attempting to decipher the mass spectrum of
accreted substructures, we currently lack a clear picture of what made up the stellar halo, and when. In this letter, we use an alternative
approach to gain insight into the progenitors of the Galactic
halo. Recent work by \cite{momany14} (see also \citealt{momany07})
showed that the number ratio of blue straggler (BS) to horizontal
branch (HB) stars in dwarfs and globular clusters is dependent on the
satellite's stellar mass. Hence, this ratio could potentially be used
to constrain the mass spectrum of substructures that contributed to
the stellar halo. With this aim in mind, we provide a careful
comparison between the number ratio of BS-to-blue horizontal branch
(BHB) stars in different Milky Way companions (classical dwarfs,
ultra-faint dwarfs and globular clusters) and the stellar halo
overall.

\vspace{-10pt}
\section{A-type Star Populations in the Milky Way Halo}

In this section, we identify the BHB and BS populations in dwarf
galaxies, globular clusters and the stellar halo. \cite{momany14} (also \citealt{momany07}) showed that the ratio of BS to HB stars varies as a function of
luminosity for satellites in the Milky Way. However, in their study
the entire HB was considered, which includes the red horizontal branch
(RHB) and the extended blue tail of the HB. The RHB is notoriously difficult to identify in the stellar
halo, and current BS-to-HB ratios in the stellar halo are upper limits
as only BHB stars are included. Hence, in this work we consider the
BS-to-BHB ratio for a fair comparison between satellites and the field
halo. Our use of BHB stars on the denominator of this population ratio could be perceived as problematic, particularly if the BHB population is scarce, or does not exist at all in some satellites. However, it is worth pointing out that, to our knowledge, there isn't a single dwarf galaxy that does not have any BHB stars. On the other hand, some very metal-rich globular clusters are devoid of a BHB population, and we discuss this further in Section \ref{sec:gcs}.

We note that our choice of BS-to-BHB ratio as a probe of the stellar halo progenitors is made for both physical and practical reasons. The BS-to-BHB ratio is arguably the cleanest population relation that can be measured in both satellite galaxies \textit{and} the stellar halo (see Section \ref{sec:stellar_halo}). In particular, redder populations such as RHB and red giant branch (RGB) stars suffer from severe foreground contamination, and are much more difficult to isolate in the halo with photometry alone. However, the main advantage of using these A-type star populations is that the BS-to-BHB \textit{ratio} is easier to quantify in the stellar halo than the total number of BHB, BS, RGB, RHB etc. stars alone (see \citealt{deason11} and Section \ref{sec:stellar_halo}).

\vspace{-10pt}
\subsection{Dwarf Galaxies}

\begin{table*}
\centering
\renewcommand{\tabcolsep}{0.3cm}
\renewcommand{\arraystretch}{0.1}
\begin{tabular}{|l c c c c r c|}\hline
\textbf{Name} & $\mathbf{M_V}$ & \textbf{Photometry} & \textbf{FOV} & $\mathbf{N_{\rm BS}/N_{\rm BHB}}$ & \textbf{Ref} & $\mathbf{\langle N_{\rm BS}/N_{\rm BHB}\rangle}$\\
\hline
Bo\"{o}tes I & -6.3 & Blanco/Mosaic-II (g, i) & $36'\times36'$  & $1.3\pm0.4$ & B06 & $1.5\pm0.5$\\
& & Subaru/Suprime-Cam (V, I) & $34'\times27'$ &  $2.5\pm0.9$& 012 &\\
& & HST/ACS (F606W, F814W) & $(5)$ $3.4'$ $^2$&  $3.0\pm2.5$& W15&\\
\hline
Canes Venatici II & -4.9 & Subaru/Suprime-Cam (g$'$, i$'$) & $34'\times27'$  &$0.6\pm0.3$ & B07 & $0.6\pm0.2$\\
& & HST/ACS (F606W, F814W) & $3.4'$ $^2$  &$0.7\pm0.4$& W15&\\
& & HST/WFPC2 (F606W, F814W) &$2.4'$ $^2$ & $0.7\pm0.6$& H06&\\
\hline
Cetus & -11.2 &  HST/ACS (F475W, F814W) & $3.4'$ $^2$& $45.4\pm11.5$& M12& $45.4\pm11.5$\\
\hline
Coma Berenices & -4.1 & Subaru/Suprime-Cam (g$'$, i$'$) &$34'\times27'$ &$0.5\pm0.4$ & B07 &$0.5\pm0.4$\\
\hline
Draco & -8.8 & INT/WFC (V, I) & $\sim$ 1 deg$^2$ &$5.4\pm1.2$ & A01 & $5.3\pm1.1$\\
& & HST/ACS (F555W, F814W) & $3.4'$ $^2$  & $4.0\pm3.2$ & W15 &\\
\hline
Hercules & -6.6 & HST/ACS (F606W, F814W) & $3.4'$ $^2$ & $1.0\pm0.6$ & W15 & $1.0\pm0.6$\\
\hline
Leo II & -9.8 & HST/WFPC2 (F555W, F814W) & $2.4'$ $^2$  &$9.2\pm2.7$ & H05 & $10.0\pm1.9$\\
& & HST/ACS (F555W, F814W) & $3.4'$ $^2$ &$14.8\pm4.9$& W15 &\\
& & HST/WFPC2 (F606W, F814W) & $2.4'$ $^2$ &  $9.4\pm2.7$& H06 &\\
\hline
Leo IV & -5.8 &  Subaru/Suprime-Cam (V, I) & $34'\times27'$  &$3.2\pm1.5$ & 012 & $1.7\pm1.0$\\
& & HST/ACS (F606W, F814W) & $3.4'$ $^2$  & $1.0\pm1.0$& W15 &\\
\hline
Sagittarius & -13.5 & MPI/WFI (V, I) & $\sim$ 1 deg$^2$&$10.0\pm1.0$ & M03 & $10.0\pm1.0$\\
\hline
Sculptor & -11.1 & MPI/WFI (B, V, I) &  $34'\times33'$   &$2.6\pm0.1$ & R03 & $2.0\pm0.4$\\
& & CTIO/MOSAIC (V, I) & $\sim$ 4 deg$^2$ &$1.8\pm0.1$ & T11 &\\
\hline
Sextans & -9.3 & CFHT/CFH12K (B, V, I)& $42'\times28'$&$6.2\pm1.1$ & L03 &$6.2\pm1.1$\\
\hline
Tucana & -9.5 & HST/ACS (F475W, F814W) & $3.4'$ $^2$ & $3.7\pm0.3$ & M12 & $3.7\pm0.3$\\
\hline
Ursa Major I & -5.5 &  Subaru/Suprime-Cam (V, I) & $34'\times27'$ & $0.3\pm0.2$& 012 & $0.4\pm0.2$\\
& & INT/WFC (B, r) &  $23'\times12'$ &$1.5\pm1.0$ & W05 &\\
\hline
Ursa Minor & -8.8 &  INT/WFC (B, R) &  0.75 deg$^2$ &$1.0\pm0.1$ & C02 & $1.0\pm0.1$\\
& & HST/WFPC2 (F555W, F606W, F814W) & $(2)$ $2.4'$ $^2$ & $2.3\pm1.0$& H06 &\\
\hline
\end{tabular}
\caption[]{\small The dwarf galaxies used in this work. We list the dwarf name, absolute visual magnitude, the photometry used to calculate population ratios, approximate FOV, BS-to-BHB number ratio, appropriate references to the photometric data sources, and (weighted) average BS-to-BHB number ratio. A01: \cite{aparicio01}, B06: \cite{belokurov06}, B07: \cite{belokurov07}, C02: \cite{carrera02}, H05: \cite{held05}, H06: \cite{holtzman06}, L03: \cite{lee03}, M03: \cite{monaco03}, O12: \cite{okamoto12}, R03: \cite{rizzi03}, T11: \cite{deboer11}, W15: Weisz et al. in prep., W05: \cite{willman05}}
\label{tab:dwarfs}
\end{table*}

Our compilation of dwarf galaxies in the Milky Way is obtained from a
variety of photometric data sources in the literature (see Table
\ref{tab:dwarfs}). We ensure that our sample only includes photometric
data deep enough to reliably identify the BS population (typically
$\sim$2 magnitudes fainter than BHBs) from the colour-magnitude
diagram (CMD), and we only include datasets where $N_{\rm BHB} > 1$
and $N_{\rm BS} > 1$. This excludes some of the more distant dwarfs
without sufficiently deep photometry (e.g. Canes Venatici I), and some
of the ultra-faint dwarfs with very few stars (e.g. Segue I).

Our dwarf sample excludes cases with known recent star formation
(e.g. Fornax, Leo I, Carina - see e.g \citealt{weisz14}), where
contamination by young stars inhibits reliable estimates of the BS
population. Very young stars ($\sim 1-3$ Gyr, see
e.g. \citealt{santana13}) can mimic BS stars in dwarf galaxies, and we
are guided by the star formation histories derived in \cite{weisz14}
to exclude these cases where possible. For consistency, we convert all
magnitudes into SDSS bandpasses. Johnson-Cousins magnitudes are
converted to $gri$ SDSS filters using the relations in \cite{jordi06},
and HST/ACS filters are converted into Johnson-Cousins bandpasses
using the procedure outlined in \cite{sirianni05}. The magnitudes and
colours we use have been corrected for extinction following the
prescription of \cite{schlegel98}.

For a fair comparison with the stellar halo (see below), only A-type
stars with $-0.25 < g-r < 0$ are used. In cases where $g-i$ colour is
most appropriate (e.g. for $V, I$ filters), we used bright A-type
stars from SDSS ($16 < g < 17$) to calibrate a linear relation between
$g-i$ and $g-r$. We find that the colour range $-0.25 < g-r < 0.0$
roughly corresponds to $-0.44 \lesssim g -i \lesssim -0.11$ for A-type
stars. We use the globular cluster sample (see below) with $gri$
photometry to ensure that our selection of A-type stars in $g-i$ is
consistent with our selection using $g-r$.

Some example CMDs are shown in Fig. \ref{fig:cmds}. The selection
region for BHB and BS stars are shown with the blue and red polygons,
respectively. We use the Trilegal Galaxy model (\citealt{girardi05})
to estimate the foreground contamination included in our A-type star
samples. The estimated foreground in the BHB and BS CMD selection
regions is subtracted before the BS-to-BHB fractions are
computed. Note that in some cases control-fields are available, and we
use these to ensure that our estimated contamination from the Trilegal
model is doing a reasonable job. In general, the contamination in the
blue ($g-r < 0$) region of CMD space probed in this work is minimal.

The resulting BS-to-BHB ratios are given in Table \ref{tab:dwarfs} and
shown in Fig. \ref{fig:ratios}. The quoted error estimates only
include Poisson noise. For datasets where we are privy to the full photometric error distribution, we find that our measurements are not significantly affected by photometric uncertainties, and in most cases, the error budget is indeed dominated by number statistics. There are several unavoidable sources of error apparent when computing the BS-to-BHB ratio: (i) uncertain foreground/background subtraction; (ii) confusion between BS stars and normal main sequence
stars; and, (iii) radial gradients in dwarfs. However, the general
agreement between BS-to-BHB ratios from different data sources
(different FOV, filters, sample size etc.) of \textit{the same dwarf}
is encouraging, and suggests that these potential systematic
uncertainties are not significantly affecting our results. Where there
are multiple data sources for the same dwarf we show the weighted (by
inverse variance) mean value of $N_{\rm BS}/N_{\rm BHB}$ in
Fig. \ref{fig:ratios}.

\subsection{Globular Clusters}
\label{sec:gcs}
We also show in Fig. \ref{fig:ratios} the BS-to-BHB number ratio for
globular clusters in the \cite{an08} sample. These globular clusters
have SDSS photometry and the BHB and BS populations are identified
from the CMDs in the same way as the dwarf galaxies. We only include
globular clusters with $N_{\rm BHB} > 1$ and $N_{\rm BS} > 1$, and
ensure that the photometry is deep enough to identify the BS
population. This leaves a sample of 12 globular clusters that satisfy
our requirements. An example of a globular cluster CMD is shown in the
right-hand panel of Fig. \ref{fig:cmds}.

We note that we do not include the relatively metal-rich
($\mathrm{[Fe/H]} \gtrsim -0.8$) globular clusters that do not have a
BHB population, but do have BS stars (see e.g
\citealt{piotto02}). These systems would boast abnormally high
BS-to-BHB ratios and could potentially contribute BS stars to the
halo. However, given the metal-poor nature of the stellar halo ($\langle
\mathrm{[Fe/H]} \rangle \sim -1.5$, \citealt{ivezic08};
\citealt{an13}) it is reasonable to assume that these metal-rich
globular clusters are not significant contributors.

\vspace{-10pt}
\subsection{Stellar Halo}
\label{sec:stellar_halo}
\begin{figure}
  \centering
   \includegraphics[width=8.5cm, height=8.5cm]{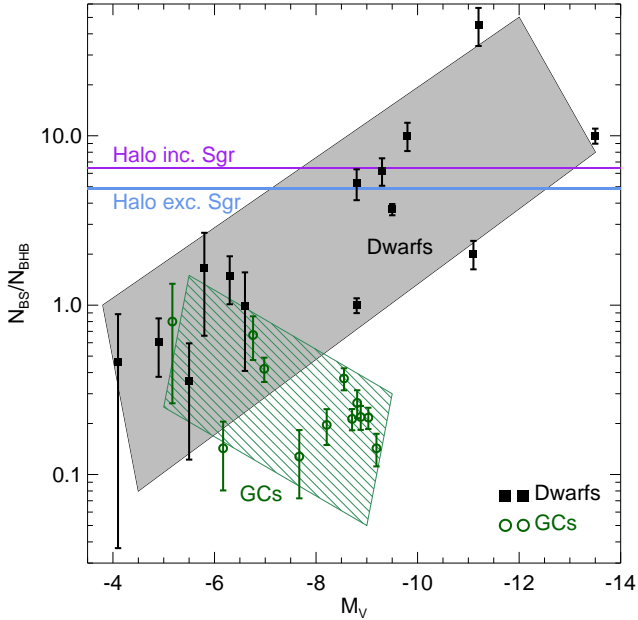}
   \caption{\small Number ratio of BS-to-BHB stars as a function of absolute visual magnitude. Dwarfs and globular clusters are shown with black squares and green circles, respectively. The quoted error bars only include Poisson noise. The halo number ratio is computed using the results of DBE11; the derived halo density profile is used to convert the overall number ratio (in a fixed magnitude slice) to a number ratio at fixed volume. The solid gray and line-filled green bands illustrate the correlation between $N_{\rm BS}/N_{\rm BHB}$ and $M_V$ for dwarfs and globular clusters, respectively.}
\vspace{-10pt}
   \label{fig:ratios}
\end{figure}

The identification of BS and BHB stars in the stellar halo is not as
straightforward. While at bright magnitudes ($g \lesssim 18.5$),
A-type stars can easily be distinguished from white dwarfs and quasars
using $ugr$ photometry, BS and BHB stars cannot be cleanly separated
using photometry alone.

In \cite{deason11} (hereafter, DBE11), we used A-type stars selected
from SDSS to measure the density profile of the stellar halo out to $D
\sim 40$ kpc. DBE11 took advantage of the overlapping, but distinct,
$ugr$ distributions of BS and BHB stars (see Fig. 2 in DBE11). The BHB
and BS populations were modeled simultaneously with class
probabilities based on $ugr$ photometry alone. This method resulted in
two quantities important for this work: 1) an estimate of the number
ratio of BS-to-BHB stars in a fixed magnitude slice (see Table 1 in
DBE11) and 2) a measure of the stellar halo density profile, under the
assumption that both BHB and BS populations follow the same density
profile.

In order to compare the stellar halo with the satellite populations,
we must take into account the different volumes probed by BHB and BS
stars in a fixed magnitude slice (BS stars are $\sim 2$ mag fainter
than BHB stars). Thus, we use the ratio $\rho^0_{\rm BS}/\rho^0_{\rm
  BHB}$, where $\rho^0_{\rm BS}=N_{\rm BS}/V_{\rm BS}$ and
$\rho^0_{\rm BHB}=N_{\rm BHB}/V_{\rm BHB}$. Here, $N_{\rm BS}$ and
$N_{\rm BHB}$ are the numbers of BS and BHB stars in a fixed magnitude
slice, and the volumes ($V_{\rm BS}, V_{\rm BHB}$) are given by
equation (9) in DBE11. The resulting ratios are $4.9 \pm 0.1$ and $6.4
\pm 0.1$ when stars belonging to the Sagittarius stream are
excluded\footnote{Using the same mask defined in DBE11.} or included,
respectively. The error estimates take into account the different
likelihoods of stellar halo density models. The halo ratios are shown
with the purple and blue lines in Fig. \ref{fig:ratios}.

\section{Population Ratios: Comparing Satellites with the Stellar Halo}
\label{sec:results}

Our compilation of BS-to-BHB number ratios for halo populations is
shown in Fig. \ref{fig:ratios} as a function of absolute
magnitude. Dwarfs and globular clusters are displayed with the solid
black squares and open green circles, respectively. The BS-to-BHB
number ratios increase with luminosity for the dwarfs, but the
opposite trend is seen for the globular clusters.

\cite{momany14} showed that the BS-to-HB ratio for dwarfs
\textit{decreases} with absolute magnitude. We find the opposite trend
for dwarf galaxies when only the BHB population is included on the
denominator. This difference is because \cite{momany14} includes
\textit{all} HB stars (BHB, RHB and the extended blue tail of the HB) in
their analysis, so their trend is likely due to the more massive
dwarfs having a more prominent RHB. As stated earlier, the RHB
population is extremely difficult to quantify in the stellar halo, so
the BS-to-BHB ratio provides a more robust comparison between
satellites and halo stars.

The difference in trends shown for GCs and dwarf galaxies is likely
related to the different BS formation mechanisms in these systems. The
two main established routes of BS production (see e.g
\citealt{davies04}), from collisional-binaries and primordial,
wide-binaries, have different significances in dwarfs and
clusters; both formation channels act in globular clusters, whereas
the low stellar density environments of dwarf galaxies precludes the
occurrence of collisional binaries. Additionally, the higher densities
(and collisional probabilities) in more massive globular clusters can
lead to BS disruption (through 3-body interactions), but this process
is not important for the (similar mass) dwarfs. This likely explains
the large differences in $N_{\rm BS}/N_{\rm BHB}$ fractions at $M_V
\sim -9$ between dwarfs and globular clusters.

The fainter dwarfs ($M_V \gtrsim -7.5$) have similar BS-to-BHB number
ratios to globular clusters at comparable luminosities, whereas more
massive dwarfs have much higher ratios than globular clusters. 

While the brighter Milky Way dwarfs have much larger BS-to-BHB number
ratios than the fainter dwarfs, there is also a good deal of
scatter. For example, Sculptor and Cetus have very similar absolute
magnitudes ($M_V \sim -11$) but very different number ratios, $N_{\rm
  BS}/N_{\rm BHB} \sim 2$ for Sculptor and $N_{\rm BS}/N_{\rm BHB}
\sim 40$ for Cetus\footnote{The unusually high BS-to-BHB number ratio
  in Cetus may result from contamination by young ($1-2$ Gyr)
  stars. Yet, to our knowledge, there is no evidence for such
  population in the dwarf.}. The star formation histories of these two
dwarfs derived by \cite{weisz14} from \textit{HST} photometry are also
very different, where Sculptor has a much older stellar population. It
is clear that at fixed luminosity the star formation histories (and
hence BS-to-BHB number ratios) can vary substantially, especially for
more massive dwarfs.

Despite the large scatter for bright dwarfs, it is clear that the low
BS-to-BHB number ratios for ultra-faint dwarfs ($N_{\rm BS}/N_{\rm
  BHB} \sim 1$) and globular clusters ($N_{\rm BS}/N_{\rm BHB} < 1$)
\textit{are not compatible with the relatively high BS-to-BHB ratio in
  the Milky Way stellar halo}. Thus, it is unlikely that the bulk of
the stellar halo was built up from (a very large number of) low
luminosity systems such as ultra-faint dwarfs and/or globular
clusters. This is in agreement with the current model predictions from
$\Lambda$CDM simulations, postulating that stellar halos are generally
dominated by massive accretion events (\citealt{bullock05};
\citealt{cooper10}; \citealt{deason13}). We do note, however, that if the progenitor satellites were drastically different to the surviving populations today, then we must be more circumspect regarding our comparison with halo stars. For example, globular clusters destroyed a long time ago ($\sim 10$ Gyr) may not have had time for collisional processes to occur, and thus the BS population may be very different in these proto-clusters. On the other hand, recent work by \cite{brown12} arguing that the UF-dwarfs are predominantly ancient ($\sim 12-14$ Gyr) populations, suggests that we are not significantly biased when comparing with the ``survivors'' at these low mass-scales.

\vspace{-10pt}
\section{Conclusions}

In this Letter, we compiled a sample of BS and BHB stars in dwarf
galaxies, globular clusters and the Milky Way stellar halo with the
aim of comparing the BS-to-BHB ratio for different halo
populations. We ensure that our selection of BS and BHB stars is as
consistent as possible (i.e. using the same photometric system and
colour cuts) between different datasets, and correct the approximate
number ratio of BS-to-BHB stars in the stellar halo (at fixed
magnitude slice) for volume effects. Our main conclusions are as
follows:

\begin{itemize}

\item The number ratio of BS-to-BHB stars in dwarf galaxies increases with increasing luminosity. Ultra-faint dwarfs have $N_{\rm BS}/N_{\rm BHB} \sim 1$, while more massive dwarfs can range from $N_{\rm BS}/N_{\rm BHB} \sim 2$ to $N_{\rm BS}/N_{\rm BHB} \sim 40$. The large scatter for more massive dwarfs is probably due to the wide variation in star formation histories.

\item GCs tend to have low BS-to-BHB ratios, $N_{\rm BS}/N_{\rm BHB}
  \lesssim 1$, which \textit{decreases} with increasing luminosity. The
  different trends shown by globular clusters and dwarfs likely
  reflect the different formation mechanisms of BS stars in these two
  populations (see e.g. \citealt{santana13}; \citealt{momany14}).

\item The relatively high BS-to-BHB ratio in the stellar halo ($N_{\rm
  BS}/N_{\rm BHB} \sim 5-6$) is inconsistent with the low ratios found
  for ultra-faint dwarfs and globular clusters. This result argues
  against ultra-faints and globular clusters being the dominant
  ``building blocks'' of the stellar halo, and instead favours more
  massive dwarfs as the more predominant progenitors.

\end{itemize}

\vspace{-10pt}
\section*{Acknowledgments}
We thank Yazan Momany, Mateo Monelli, Sakurako Okamoto and Myung Goon Lee for kindly sharing their photometric data with us. We also thank Francesca De Angeli, Yazan Momany and an anonymous referee for providing useful comments on the paper. AJD and DRW are currently supported by NASA through Hubble Fellowship grants HST-HF-51302.01,51331.01, awarded by the Space Telescope Science Institute, which is operated by the Association of Universities for Research in Astronomy, Inc., for NASA, under contract NAS5-26555. We thank the Aspen Center for Physics and the NSF Grant \#1066293 for hospitality during the conception of this paper. The research leading to these results has received funding from the European Research Council under the European Union's Seventh Framework Programme (FP/2007-2013) / ERC Grant Agreement n. 308024.

\label{lastpage}


\begin{thebibliography}{}

\bibitem[\protect\citeauthoryear{{An} et~al.,}{{An}  et~al.}{2008}]{an08}
{An} D.,  et~al., 2008, \apjs, 179, 326

\bibitem[\protect\citeauthoryear{{An} et~al.,}{{An}  et~al.}{2013}]{an13}
{An} D.,  et~al., 2013, \apj, 763, 65

\bibitem[\protect\citeauthoryear{{Aparicio}, {Carrera} \&
  {Mart{\'{\i}}nez-Delgado}}{{Aparicio} et~al.}{2001}]{aparicio01}
{Aparicio} A.,  {Carrera} R.,    {Mart{\'{\i}}nez-Delgado} D.,  2001, \aj, 122,
  2524

\bibitem[\protect\citeauthoryear{{Belokurov} et~al.,}{{Belokurov}
  et~al.}{2006}]{belokurov06}
{Belokurov} V.,  et~al., 2006, \apjl, 647, L111

\bibitem[\protect\citeauthoryear{{Belokurov} et~al.,}{{Belokurov}
  et~al.}{2007}]{belokurov07}
{Belokurov} V.,  et~al., 2007, \apj, 654, 897

\bibitem[\protect\citeauthoryear{{Brown} et~al.,}{{Brown}
  et~al.}{2012}]{brown12}
{Brown} T.~M.,  et~al., 2012, \apjl, 753, L21

\bibitem[\protect\citeauthoryear{{Bullock} \& {Johnston}}{{Bullock} \&
  {Johnston}}{2005}]{bullock05}
{Bullock} J.~S.,  {Johnston} K.~V.,  2005, \apj, 635, 931

\bibitem[\protect\citeauthoryear{{Carrera}, {Aparicio},
  {Mart{\'{\i}}nez-Delgado} \& {Alonso-Garc{\'{\i}}a}}{{Carrera}
  et~al.}{2002}]{carrera02}
{Carrera} R.,  {Aparicio} A.,  {Mart{\'{\i}}nez-Delgado} D.,
  {Alonso-Garc{\'{\i}}a} J.,  2002, \aj, 123, 3199

\bibitem[\protect\citeauthoryear{{Carretta}, {Bragaglia}, {Gratton},
  {Recio-Blanco}, {Lucatello}, {D'Orazi} \& {Cassisi}}{{Carretta}
  et~al.}{2010}]{carretta10}
{Carretta} E.,  {Bragaglia} A.,  {Gratton} R.~G.,  {Recio-Blanco} A.,
  {Lucatello} S.,  {D'Orazi} V.,    {Cassisi} S.,  2010, \aap, 516, A55

\bibitem[\protect\citeauthoryear{{Clementini}}{{Clementini}}{2010}]{clementini10}
{Clementini}, G. 2010, in Variable Stars, the Galactic halo and Galaxy Formation, eds. C. Sterken, N. Samus, \& L. Szabados (Moscow: Sternberg Astronomical
Institute of Moscow Univ.), 107

\bibitem[\protect\citeauthoryear{{Cooper} et~al.,}{{Cooper}
  et~al.}{2010}]{cooper10}
{Cooper} A.~P.,  et~al., 2010, \mnras, 406, 744

\bibitem[\protect\citeauthoryear{{Davies}, {Piotto} \& {de Angeli}}{{Davies}
  et~al.}{2004}]{davies04}
{Davies} M.~B.,  {Piotto} G.,    {de Angeli} F.,  2004, \mnras, 349, 129

\bibitem[\protect\citeauthoryear{{Deason}, {Belokurov} \& {Evans}}{{Deason}
  et~al.}{2011}]{deason11}
{Deason} A.~J.,  {Belokurov} V.,    {Evans} N.~W.,  2011, \mnras, 416, 2903

\bibitem[\protect\citeauthoryear{{Deason}, {Belokurov}, {Evans} \&
  {Johnston}}{{Deason} et~al.}{2013}]{deason13}
{Deason} A.~J.,  {Belokurov} V.,  {Evans} N.~W.,    {Johnston} K.~V.,  2013,
  \apj, 763, 113

\bibitem[\protect\citeauthoryear{{de Boer} et~al.,}{{de Boer}
  et~al.}{2011}]{deboer11}
{de Boer} T.~J.~L., et~al., 2011, \aap, 528, A119

\bibitem[\protect\citeauthoryear{{Fiorentino} et~al.,}{{Fiorentino}
  et~al.}{2014}]{fiorentino14}
{Fiorentino}, G., et~al., 2014, ApJL in press, arXiv:1411.7300

\bibitem[\protect\citeauthoryear{{Font}, {Johnston}, {Bullock} \&
  {Robertson}}{{Font} et~al.}{2006}]{font06}
{Font} A.~S.,  {Johnston} K.~V.,  {Bullock} J.~S.,    {Robertson} B.~E.,  2006,
  \apj, 646, 886

\bibitem[\protect\citeauthoryear{{Frebel}, {Simon}, {Geha} \&
  {Willman}}{{Frebel} et~al.}{2010}]{frebel10}
{Frebel} A.,  {Simon} J.~D.,  {Geha} M.,    {Willman} B.,  2010, \apj, 708, 560

\bibitem[\protect\citeauthoryear{{Girardi}, {Groenewegen}, {Hatziminaoglou} \&
  {da Costa}}{{Girardi} et~al.}{2005}]{girardi05}
{Girardi} L.,  {Groenewegen} M.~A.~T.,  {Hatziminaoglou} E.,    {da Costa} L.,
  2005, \aap, 436, 895

\bibitem[\protect\citeauthoryear{{Held}}{{Held}}{2005}]{held05}
{Held}, E.~V., 2005, in IAU Colloq. 198: Near-field cosmology with dwarf elliptical galaxies, eds. {Jerjen} H.,  {Binggeli} B., 11

\bibitem[\protect\citeauthoryear{{Holtzman}, {Afonso} \& {Dolphin}}{{Holtzman}
  et~al.}{2006}]{holtzman06}
{Holtzman} J.~A.,  {Afonso} C.,    {Dolphin} A.,  2006, \apjs, 166, 534

\bibitem[\protect\citeauthoryear{{Ivezi{\'c}} et~al.,}{{Ivezi{\'c}}
  et~al.}{2008}]{ivezic08}
{Ivezi{\'c}} {\v Z}.,  et~al., 2008, \apj, 684, 287

\bibitem[\protect\citeauthoryear{{Jordi}, {Grebel} \& {Ammon}}{{Jordi}
  et~al.}{2006}]{jordi06}
{Jordi} K.,  {Grebel} E.~K.,    {Ammon} K.,  2006, \aap, 460, 339

\bibitem[\protect\citeauthoryear{{Lee} et~al.,}{{Lee}  et~al.}{2003}]{lee03}
{Lee} M.~G.,  et~al., 2003, \aj, 126, 2840

\bibitem[\protect\citeauthoryear{{Martell}, {Smolinski}, {Beers} \&
  {Grebel}}{{Martell} et~al.}{2011}]{martell11}
{Martell} S.~L.,  {Smolinski} J.~P.,  {Beers} T.~C.,    {Grebel} E.~K.,  2011,
  \aap, 534, A136

\bibitem[\protect\citeauthoryear{{Momany}, {Held}, {Saviane}, {Zaggia}, {Rizzi}
  \& {Gullieuszik}}{{Momany} et~al.}{2007}]{momany07}
{Momany} Y.,  {Held} E.~V.,  {Saviane} I.,  {Zaggia} S.,  {Rizzi} L.,
  {Gullieuszik} M.,  2007, \aap, 468, 973

\bibitem[\protect\citeauthoryear{{Momany}}{{Momany}}{2014}]{momany14}
{Momany} Y.,  2014, in Ecology of Blue Straggler Stars (Ch6), eds. H.M.J. Boffin, G. Carraro \& G. Beccari, Astrophysics and Space Science Library, Springer

\bibitem[\protect\citeauthoryear{{Monaco}, {Bellazzini}, {Ferraro} \&
  {Pancino}}{{Monaco} et~al.}{2003}]{monaco03}
{Monaco} L.,  {Bellazzini} M.,  {Ferraro} F.~R.,    {Pancino} E.,  2003, \apjl,
  597, L25

\bibitem[\protect\citeauthoryear{{Okamoto}, {Arimoto}, {Yamada} \&
  {Onodera}}{{Okamoto} et~al.}{2012}]{okamoto12}
{Okamoto} S.,  {Arimoto} N.,  {Yamada} Y.,    {Onodera} M.,  2012, \apj, 744,
  96

\bibitem[\protect\citeauthoryear{{Oosterhoff}}{{Oosterhoff}}{1939}]{oosterhoff39}
{Oosterhoff} P.~T.,  1939, The Observatory, 62, 104

\bibitem[\protect\citeauthoryear{{Piotto} et~al.,}{{Piotto}
  et~al.}{2002}]{piotto02}
{Piotto} G.,  et~al., 2002, \aap, 391, 945

\bibitem[\protect\citeauthoryear{{Rizzi}, {Held}, {Momany}, {Saviane},
  {Bertelli} \& {Moretti}}{{Rizzi} et~al.}{2003}]{rizzi03}
{Rizzi} L.,  {Held} E.~V.,  {Momany} Y.,  {Saviane} I.,  {Bertelli} G.,
  {Moretti} A.,  2003, \memsai, 74, 510

\bibitem[\protect\citeauthoryear{{Robertson}, {Bullock}, {Font}, {Johnston} \&
  {Hernquist}}{{Robertson} et~al.}{2005}]{robertson05}
{Robertson} B.,  {Bullock} J.~S.,  {Font} A.~S.,  {Johnston} K.~V.,
  {Hernquist} L.,  2005, \apj, 632, 872

\bibitem[\protect\citeauthoryear{{Santana}, {Mu{\~n}oz}, {Geha},
  {C{\^o}t{\'e}}, {Stetson}, {Simon} \& {Djorgovski}}{{Santana}
  et~al.}{2013}]{santana13}
{Santana} F.~A.,  {Mu{\~n}oz} R.~R.,  {Geha} M.,  {C{\^o}t{\'e}} P.,  {Stetson}
  P.,  {Simon} J.~D.,    {Djorgovski} S.~G.,  2013, \apj, 774, 106

\bibitem[\protect\citeauthoryear{{Schlegel}, {Finkbeiner} \&
  {Davis}}{{Schlegel} et~al.}{1998}]{schlegel98}
{Schlegel} D.~J.,  {Finkbeiner} D.~P.,    {Davis} M.,  1998, \apj, 500, 525

\bibitem[\protect\citeauthoryear{{Sesar}, {Juri{\'c}} \& {Ivezi{\'c}}}{{Sesar}
  et~al.}{2011}]{sesar11}
{Sesar} B.,  {Juri{\'c}} M.,    {Ivezi{\'c}} {\v Z}.,  2011, \apj, 731, 4

\bibitem[\protect\citeauthoryear{{Sirianni} et~al.,}{{Sirianni}
  et~al.}{2005}]{sirianni05}
{Sirianni} M.,  et~al., 2005, \pasp, 117, 1049

\bibitem[\protect\citeauthoryear{{Tolstoy}, {Venn}, {Shetrone}, {Primas},
  {Hill}, {Kaufer} \& {Szeifert}}{{Tolstoy} et~al.}{2003}]{tolstoy03}
{Tolstoy} E.,  {Venn} K.~A.,  {Shetrone} M.,  {Primas} F.,  {Hill} V.,
  {Kaufer} A.,    {Szeifert} T.,  2003, \aj, 125, 707

\bibitem[\protect\citeauthoryear{{Venn}, {Irwin}, {Shetrone}, {Tout}, {Hill} \&
  {Tolstoy}}{{Venn} et~al.}{2004}]{venn04}
{Venn} K.~A.,  {Irwin} M.,  {Shetrone} M.~D.,  {Tout} C.~A.,  {Hill} V.,
  {Tolstoy} E.,  2004, \aj, 128, 1177

\bibitem[\protect\citeauthoryear{{Weisz}, {Dolphin}, {Skillman}, {Holtzman},
  {Gilbert}, {Dalcanton} \& {Williams}}{{Weisz} et~al.}{2014}]{weisz14}
{Weisz} D.~R.,  {Dolphin} A.~E.,  {Skillman} E.~D.,  {Holtzman} J.,  {Gilbert}
  K.~M.,  {Dalcanton} J.~J.,    {Williams} B.~F.,  2014, \apj, 789, 147

\bibitem[\protect\citeauthoryear{{Willman} et~al.,}{{Willman}
  et~al.}{2005}]{willman05}
{Willman} B.,  et~al., 2005, \apjl, 626, L85

\end{thebibliography}
\end{document}